\documentstyle[12pt]{article}

\def\?#1{#1}
\?{
\textwidth 6truein \textheight 8.5truein \oddsidemargin 0.25truein
\evensidemargin 0.25truein \topmargin -0.3truein \headsep 0.4truein
\parskip 2ptplus2pt
}

\newif\iffigs \figsfalse 

\def\ds{\displaystyle} \def\rset{{\rm I\kern -0.2em R}}
\def\eql{\,{\stackrel{\rm law}{=}}\,}
\def\IR{\hbox{\rm I\kern-.2em\hbox{\rm R}}}
\def\proof{\smallskip \noindent {\bf Proof. \ }}
\newcommand\filledsquare{\ \vrule width 1.5ex height 1.2ex}  
\def\qed{\hfill\filledsquare\linebreak\smallskip\par}
\newcommand\Lpar{\left(}                \newcommand\Rpar{\right)}
\newtheorem{lemma}{Lemma}[section]       
\newtheorem{remark}[lemma]{Remark}            

   \makeatletter \def\fps@figure{tp} \makeatother
\iffigs
\fi

\def\drawing #1 #2 #3 {\begin{center} \setlength{\unitlength}{1mm}
                   \begin{picture}(#1,#2)(0,0) 
                   \put(0,0){\framebox(#1,#2){#3}} \end{picture} \end{center}}

\begin{document}
\centerline{\LARGE Bifractality of the Devil's staircase appearing in}
\vspace{1mm}
\centerline{\LARGE the Burgers equation with Brownian initial velocity}
\vspace{8mm}
\centerline{\large E.~Aurell,$^{1,2}$ U.~Frisch,$^3$ A.~Noullez$^3$
            and M.~Blank$^{3,4}$}
\vspace{2mm}
\centerline{$^1$ Center for Parallel Computers, Royal Institute of 
            Technology, S--100 44 Stockholm, Sweden}
\centerline{$^2$ Mathematics Dept., Stockholm University, S--106 91 
            Stockholm, Sweden}
\centerline{$^3$ Observatoire de Nice, URA CNRS 1362, B.P. 4229,
            06304 Nice Cedex 4, France}
\centerline{$^4$ Institute for Information Transmission Problems,
            B.~Karetnij Per. 19, 101477 Moscow, Russia}
\vspace{2mm}

\centerline{November 13, 1996}
\vspace{3mm}
\centerline{Submitted to \sl J. Stat. Phys.}
\vspace{5mm}
\begin{abstract}
It is shown that the inverse Lagrangian map for the solution of the
Burgers equation (in the inviscid limit) with Brownian initial
velocity presents a bifractality (phase transition) similar to that of
the Devil's staircase for the standard triadic Cantor set. Both
heuristic and rigorous derivations are given. It is explained why
artifacts can easily mask this phenomenon in numerical simulations.
\end{abstract}

\section{Introduction}

There is a renewed  interest among physicists and mathematicians in
the Burgers equation 
\begin{equation}
\partial_t v  +v\partial_x v= \nu \partial_x^2 v.
\label{burgers}
\end{equation}
In particular it has been discovered that, when the initial velocity
$v_0(x)$ is a Brownian (or fractional Brownian) motion function of the
space coordinate $x$ and the limit of vanishing viscosity $\nu$ is
taken, the Lagrangian map 
\begin{equation}
a \mapsto x(a,t)
\label{defLmap}
\end{equation}
from the initial fluid particle position $a$
to its position $x$ at time $t$ is a Devil's staircase
\cite{sinai,SAF,verg}; this has consequences for the distributions of
masses of large-scale structures in the Universe \cite{verg}. 

Since the Lagrangian map (\ref{defLmap}) is nondecreasing (a
consequence of the fact that fluid particles can merge but not cross),
two nonnegative measures may be defined via their increments. The
direct Lagrangian measure associates to a Lagrangian interval $[a,b]$
the length $\Delta x =x(b)-x(a)$. The inverse Lagrangian measure
associates to an Eulerian interval $[x,y]$ the length $\Delta a=
a(y)-a(x)$. When the initial density field is uniform, this length
$\Delta a$ is proportional to the mass in the interval and will
therefore be called the ``mass''. The multifractal properties of the
direct and inverse Lagrangian measures can be analyzed by studying the
scaling properties of the moments of their increments. Here, we shall be
interested only in the inverse Lagrangian measure, whose moments are
defined by
\begin{equation}
M_q(l) \equiv  \langle [a(x+l) - a(x)]^q\rangle,\qquad q\ge 0,
\label{defMql}
\end{equation}
which does not depend on $x$ because the Lagrangian map has
homogeneous (stationary in the space variable) increments.

In Ref.~\cite{SAF} it was conjectured that the inverse Lagrangian
measure has a bifractality similar to that known for the standard Devil's
staircase associated to the triadic Cantor set (see
Fig.~\ref{f:Devil}).  According to the conjecture, the
scaling exponents $\tau_q$, obtained from the small-$l$ behavior of
the moments
\begin{equation}
M_q(l) \propto l^{\tau_q},
\label{deftauq}
\end{equation}
should present a {\em phase transition}\,: for $q\ge q_\star$, one
should have $\tau_q=1$, while for $0\le q\le q_\star$, one should have
$\tau_q = q/q_\star$. Phase transitions of this sort are frequently
observed in studying fractal sets of physical interest (see, e.g.,
Ref.\cite{ACK}). Preliminary numerical tests, reported in
Refs.~\cite{SAF,verg} were rather inconclusive.  Further (unpublished)
simulations indicated the presence of a small-$l$ scaling r\'egime
with $\tau_q=1$ for any $q$. While developing the correct theory for
the inverse Lagrangian map, we found that this is actually an artifact
inherent to discrete numerical simulations which can hide the true
scaling.

Here, we demonstrate the conjectured bifractality.
Section~\ref{s:formulation} is devoted to recalling some known results
about the solution of the Burgers equation and to formulating the
problem with (fractional) Brownian initial conditions. In
Section~\ref{s:heuristic} we present a heuristic approach (the
physicist's viewpoint); the arguments are not rigorous, but encompass
both the Brownian and the fractional Brownian case; for pedagogical
reasons, the arguments are presented first for the standard Devil's
staircase. In Section~\ref{s:rigorous}, we present a rigorous proof
for the Brownian case. In Section~\ref{s:simulations}, we present
numerical simulations, discuss the aforementioned artifact and show
how to overcome it.

\section{The Lagrangian map for the Burgers equation}
\label{s:formulation}

We recall the construction of the solution to
(\ref{burgers}) (see Refs.~\cite{sinai,SAF,verg} for details). The
solution at time $t$,  for a continuous initial velocity $v_0(x)$,
is given by
\begin{equation}
v(x,t)= v_0(a(x,t)).
\label{lagsol}
\end{equation}
Here, $x\mapsto a(x,t)$ is the inverse Lagrangian map obtained as
follows\,: we denote by $\psi_0 (x)$ the initial  velocity potential
($v(x)=-\partial_x \psi(x)$); then the Lagrangian location $a(x,t)$ is such
that $\psi_0 (a) - (x-a)^2/(2t)$ achieves its global maximum.  This map is
nondecreasing and has discontinuities at shock locations. Its inverse
$a \mapsto x(a,t)$ is called the Lagrangian map. Those Lagrangian
locations which are not mapped into shocks are called regular.

For the case of fractional Brownian initial velocity,  
$v_0(x)$ is a random function,  defined from
$-\infty$ to $+\infty$, which is Gaussian  and satisfies the following 
conditions (angular brackets denote averages)\,: 
\begin{eqnarray}
\langle v_0(x) \rangle &=& 0, \label{vzeroavzer}\\
\langle \left[v_0(x')-v_0(x)\right]^2\rangle &=& C^2|x'-x|^{2h},
\label{structfunc}
\end{eqnarray}
with $0<h<1$. The case $h=1/2$ is the standard Brownian motion curve
(in the space variable) for which it was shown in Ref.~\cite{sinai}
that the regular points form (almost surely) a set of Hausdorff
dimension $D=1/2$. For the case $h\ne 1/2$, very strong numerical
evidence was given in Ref.~\cite{verg} that $D=h$. It follows that 
the Lagrangian map is a Devil's staircase. 

The scaling properties of the fractional Brownian motion $v_0(x)$
and of the Burgers equation imply that \cite{SAF,verg}
\begin{equation}
v(x,t)  \eql t^{h\over1-h} v\left(x t^{-{1\over1-h}}, 1\right),
\label{rescu}
\end{equation}
where $\eql$ denotes probabilistic equality in law.  Hence, the
knowledge of the statistical properties at time $t=1$ gives those at
any other $t>0$ by simple rescaling. Henceforth, in discussing the
theory, we shall sometimes set $t=1$ without loss of generality. In
performing numerical simulations with a finite mesh and a maximum box
size $L$, the choice of the time $t$ becomes relevant since the only
scales which can be meaningfully related to the untruncated problem
are between the mesh-size and $L$.\footnote{Or a subset thereof, as we
shall see in Section~\ref{s:simulations}.} By changing $t$ we can
adjust the coalescence length 
\begin{equation}
l_c(t)\equiv \left(Ct\right)^{1/(1-h)}
\label{defcoal}
\end{equation}
so that it does not lie too close to either of these.
Over separations $l \gg l_c(t)$ the velocity increments, as well as
the Lagrangian measures, remain basically
unaffected, because not enough time has elapsed for substantial
particle merging. In other words, the Lagrangian map is close to the
identity.

\section{A heuristic approach}
\label{s:heuristic}

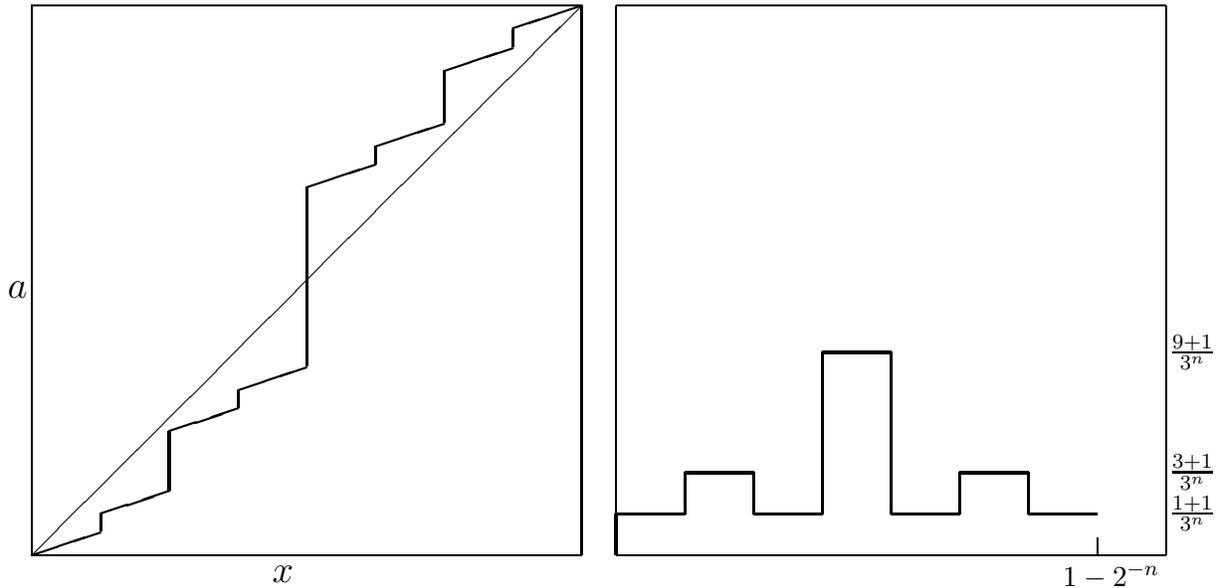
\begin{figure} \begin{center} \setlength{\unitlength}{1.3pt}
\begin{picture}(350,160){\large
       \put(0,0){\line(1,0){160}}   \put(0,0){\line(0,1){160}}
       \put(0,160){\line(1,0){160}} \put(160,0){\line(0,1){160}}
       \put(0,0){\line(1,1){160}}
       \thicklines
       \put(0,0){\line(3,1){20}}     \put(20,7){\line(0,6){5}}
       \put(20,12){\line(3,1){20}}   \put(40,19){\line(0,6){17}}
       \put(40,36){\line(3,1){20}}   \put(60,43){\line(0,6){5}}
       \put(60,48){\line(3,1){20}}   \put(80,55){\line(0,6){52}}   
       \put(80,107){\line(3,1){20}}  \put(100,114){\line(0,5){5}}
       \put(100,119){\line(3,1){20}} \put(120,126){\line(0,5){15}}
       \put(120,141){\line(3,1){20}} \put(140,148){\line(0,5){5}}
       \put(160,160){\line(-3,-1){20}} \put(70,-8){$x$} \put(-7,75){$a$}
       \put(170,0){\begin{picture}(160,160) \thinlines
                   \put(0,0){\line(1,0){160}}   \put(0,0){\line(0,1){160}}
                   \put(0,160){\line(1,0){160}} \put(160,0){\line(0,1){160}}
                   \put(140,0){\line(0,1){5}}
                   \thicklines
                   \put(0,0){\line(0,1){12}}    \put(0,12){\line(1,0){20}}
                   \put(20,12){\line(0,1){12}}  \put(20,24){\line(1,0){20}}
                   \put(40,24){\line(0,-1){12}} \put(40,12){\line(1,0){20}}
                   \put(60,12){\line(0,1){47}}  \put(60,59){\line(1,0){20}}
                   \put(80,59){\line(0,-1){47}} \put(80,12){\line(1,0){20}}
                   \put(100,12){\line(0,1){12}} \put(100,24){\line(1,0){20}}
                   \put(120,24){\line(0,-1){12}}\put(120,12){\line(1,0){20}}
                   \put(161,10){\small$\frac{1+1}{3^n}$}
                   \put(161,22){\small$\frac{3+1}{3^n}$}
                   \put(161,57){\small$\frac{9+1}{3^n}$}
                   \put(130,-9){\small$1-2^{-n}$}
                  \end{picture}}
}\end{picture} \end{center}
\caption{Inverse Devil's staircase and the corresponding increment function}
\label{f:Devil} \end{figure}

We begin by recalling in some detail how bifractality
arises  for the standard inverse Devil's staircase (Fig.~\ref{f:Devil}).
Let $a$ denote the coordinate on the vertical (Lagrangian)
axis and $x$ the coordinate on the horizontal (Eulerian) axis. At
all (dyadic) $x$'s which are integer multiples of $2^{-n}$
there is a jump (shock) of the inverse Lagrangian map $a(x)$.
It is straightforward to show that there are 
\begin{equation}
\begin{array}{c}
\ds  \hbox{$2^0$jumps of amplitude $3^{-1}$}\\[1.6ex]
\ds\hbox{$2^1$ jumps of amplitude $3^{-2}$}\\[1.6ex]
\ds \ldots\\[1.6ex]
\ds \hbox{$2^n$ jumps of amplitude $3^{-(n+1)}$}\\[1.6ex]
\ds\ldots
\end{array}
\label{mdistr}
\end{equation}
(This formula, which  gives the distribution of the shock amplitudes, will
be referred to as the mass distribution.)

What is the equivalent of the quantity $M_q(l)$ defined
above\,? First, we shall restrict ourselves to increments $l$
between two successive dyadic points $x_{n,p}=p\,2^{-n}$ and
$x_{n,p+1}$ of the same $n$th generation, so that $l =
2^{-n}$. The average $\langle Q\rangle$ of a quantity $Q$ will
be just a sum over the $2^n$ intervals of this type, divided by
$2^n$. For example,  
\begin{equation}
M_q(l) = \langle[a(x+l)-a(x)]^q\rangle
\label{defmomomentsIL}
\end{equation} 
will be evaluated for $l = 2^{-n}$ as
\begin{equation}
M_q(l) =(1/2^n)\sum_{p=0}^{p=2^n-1} [a(x_{n,p+1}) -
a(x_{n,p})]^q.
\label{howeval}
\end{equation}

Now, an important {\em remark}\,: If we do not restrict the Eulerian
positions to dyadic points $x_{n,p}$, or if we work with a randomized
version of the Devil's staircase, the Lagrangian increment
$a(x') - a(x)$ will be a sum over all the Lagrangian
(shock) intervals corresponding to the dyadic
Eulerian points between $x$ and $x'$. This will generally
include  infinitely many intervals of generations $n' \ge n$.
However, since the number of such intervals grows as 
$2^{n'}$, while their length decreases as $3^{-n'}$, the sum is
dominated by just the first few terms for which $n'$ is equal
to $n$, or just a bit larger. Hence, we can replace the
Lagrangian increment over an Eulerian distance $l = 2^{-n}$
by just the contribution coming from the $n$th generation,
thereby committing an error which just affects constants but
not scaling. We thus have
\begin{equation}
M_q(l) \sim (1/2^n)\sum_{p=0}^{p=2^n-1} [\hbox{length of
$p$th Lagrangian interval of $n$th generation}]^q.
\label{stopsum}
\end{equation}

The mass distribution gives us the number of shock intervals
of a given length.  Summing over all the generations
$m \le n$, we obtain : 
\begin{equation}
M_q(l) \sim (1/2^n) 
\sum_{m=0}^{m=n}2^m [3^{-(m+1)}]^q.
\label{thesum}
\end{equation}

It is clear that we must distinguish two cases :
\par\noindent (i) when $q > D = \ln 2/\ln 3$, the sum is
dominated by its first term and we obtain
\begin{equation}
M_q(l) \sim (1/2^n) \sim l^1,
\label{case}
\end{equation}
so that $\tau_q=1$;
\par\noindent (ii) when $0 \le q < D$, the sum is dominated by
its last term and we obtain
\begin{equation}
M_q(l) \sim (1/2^n) 2^n [3^{-(n+1)}]^q \sim 3^{-nq} \sim
l^{q\ln 3/\ln 2},
\label{caseii}
\end{equation}
so that $\tau_q = q\ln 3/\ln 2$, which establishes the bifractality
for the standard Devil's staircase. This is seen as a rather
elementary result, obtained without recourse to numerical
computations, unlike the much more intricate phase transitions
encountered in the study of certain circle maps \cite{ACK}.\\

We now turn to the Burgers problem for arbitrary $0<h<1$. The mass
function is known \cite{sinai,verg}\,: the mean number per unit length
of Lagrangian intervals with a length between $\Delta a$  and
$\Delta a/2$ is
\begin{equation}
N(\Delta a) \propto (\Delta a)^{-h}.
\label{masssinai}
\end{equation} 
Sinai's  theory  of the mass function \cite{sinai} was done for $t=1$.
It is useful to introduce the correct dependence on the time of
$N(\Delta a,t)$. This is done by nondimensionalization\,: we multiply
$N(\Delta a,t)$ by the coalescence length $l_c \sim t^{1/(1-h)}$
and we divide the length $\Delta a$ by $l_c$. We thereby obtain\,:
\begin{equation}
N(\Delta a,t) \sim (\Delta a)^{-h}/t. 
\label{masswithtime}
\end{equation}

For the standard Devil's staircase, the length of the Eulerian
interval $l$ was chosen as the inverse of the mean number (per unit
length) of Lagrangian intervals for the $n$th generation.  For the
Burgers case, taking $\Delta a = 2^{-n}$ and using
(\ref{masswithtime}), this mean number is $2^{nh}/t$. This is of
course also the mean number of shocks per unit (Eulerian) length with
a mass of the order of $\Delta a = 2^{-n}$.  Hence, the corresponding
Eulerian interval, the mean distance between two such successive
shocks, is
\begin{equation}
l \sim t\, 2^{-nh}= t(\Delta a)^h.
\label{craxi}
\end{equation}
This may also be interpreted as the length of an
Eulerian interval such that the mean number of shocks having
a Lagrangian length between $2^{-n}$ and $2^{-(n+1)}$ is order
unity.

We can now just repeat for the Burgers problem essentially the
same argument as developed for the standard Devil's staircase,
to obtain\,: 
\begin{equation}
M_q(l,t) \sim {t\over 2^{nh}}\,\sum_{m=0}^{m=n} N(2^{-m},t) [2^{-m}]^q
\sim {l\over t}\,\sum_{m=0}^{m=n} 2^{mh} [2^{-m}]^q, 
\label{berlusco}
\end{equation}
where the upper limit $n$ is given in terms of $l$ by
(\ref{craxi}).

As before, depending on whether $q > q_\star = h$ or $0 \le
q<q_\star,$ the sum is dominated by its first or its last term. Hence,
we find, in the former case, that $M_q(l,t)\propto l^1$ and, in the
latter case, that $M_q(l,t)\propto l^{q/h}$, so that $\tau_q = 1$ for
$q > h$ and $\tau_q = q/h$ for $0 \le q < h,$. This establishes the
bifractality.

\section{Bifractality of shocks: rigorous theory} \label{s:rigorous}

Let $a(x)$ be a (random) nondecreasing function defined on a finite
interval $I \subset \IR$.  Denote by $M_q(a(\cdot),l)$ the space averaged
value of the $q$-th power of the increment\,: 
\begin{equation}
M_q(a(\cdot),l) :=
\frac1{|I|-l} \int_{I-(l)} [a(x+l) - a(x)]^q \, dx,
\label{mbl1}
\end{equation}
where $I-(l)$
means that the integration is over the interval $I$ with the exception of an
interval of length $l$ at the right end of the interval $I$, where the
increment is not well defined.

Our aim is to study asymptotic properties as $l \to 0$ of the
mathematical expectation (mean value) of $M_q(a(\cdot),l)$.

In Section~\ref{s:rig1} we shall derive estimates at first for the
inverse of the standard Devil's staircase and then for a general
function of this type, assuming that we know the statistical
properties of $N_k$, the number of shocks of magnitude of order
$2^{-k}$.  In Section~\ref{s:rig2} we prove inequalities for $N_k$ for
the case of the solution of Burgers equation with a Brownian initial velocity.

\subsection{Estimate of $M_q(a(\cdot),l)$}
\label{s:rig1}

Let $a(x)$ be the inverse Devil's staircase function on the interval 
$I=[0,1]$ and let $a_n$ be its approximation constructed in the same way 
as for the $n$-th step of the construction of the standard Cantor set 
(see Fig.~\ref{f:Devil}). To calculate the functional $M_q(a(\cdot),l)$ we 
calculate at first the sequence of approximations $M_q(a_m(\cdot),2^{-n})$, 
$m \ge n$ for the functions $a_m$ and then investigate its behavior
as $n \to \infty$.

For any fixed $n \le m$ the increment function $a_m(x+2^{-n}) - a_m(x)$ 
is a piecewise constant function with $2^k$ pieces of length $2^{-n}$ 
of magnitude $3^{-k}+3^{-n}$ for $k=0,1,\dots,n$ (see Fig.~\ref{f:Devil}).
Therefore,
\begin{eqnarray}
&&M_q(a_m(\cdot),2^{-n}) := \frac1{1-2^{-n}}
   \int_0^{1-2^{-n}} [a_m(x+2^{-n}) - a_m(x)]^q \, dx \nonumber\\
 &&  = \frac1{1-2^{-n}} \sum_{k=0}^{n-1} 2^{-n} 2^k 
                      \Lpar 3^{-(k+1)} + 3^{-n} \Rpar^q \nonumber\\
  && = \frac1{1-2^{-n}} \sum_{k=0}^{n-1} 2^{-n} 2^k 3^{-(k+1)q}
   + \frac1{1-2^{-n}} \sum_{k=0}^{n-1} 2^{-n} 2^k 3^{-(k+1)q} \cdot
R_k ,
\label{truc}
\end{eqnarray}
where
\begin{equation}
R_k := (1 + 3^{-n+k+1})^q - 1 
   \le \cases{ 2q \; 3^{-n+k+1}, &if $q>1$;   \cr
                     3^{-n+k+1}, &if $q\le1$; \cr} .
\label{mbl3}
\end{equation}
The first term in (\ref{truc}) can be calculated as follows\,:
\begin{equation}
 \frac1{1-2^{-n}} \sum_{k=0}^{n-1} 2^{-n} 2^k 3^{-(k+1)q} 
   = \frac{2^{-n} 3^{-q}}{1-2^{-n}} \sum_{k=0}^{n-1} 
     \Lpar \frac2{3^q} \Rpar^k 
   = \frac{2^{-n} 3^{-q}}{1-2^{-n}}
     \frac{1 - \Lpar \frac2{3^q} \Rpar^n}{1 - \frac2{3^q}},
\label{mbl4}
\end{equation}
while the second term becomes negligible compared to the first one as
$n \to \infty$.

\bigskip

Consider now a more general function $a:I \to \IR$ of this type.
For each integer $m>0$ we construct an approximation $a_m(x)$, which
is a piecewise linear function with $N_k$ jumps of magnitude $h_k \in
[2^{-k-1}, 2^{-k})$ for $0\le k <m$.  Then the integral of the
corresponding increment function with the increment $2^{-n}$ may be
estimated from above by the integral over a piecewise constant
function with $2^k$ pieces of length $2^{-n}$ of magnitude
$2^{-k}+2^{-n}$ for $0\le k <n$. There is a similar estimate from
below (with the magnitude $2^{-k-1}+2^{-n}$). Notice
that our estimates do not depend on $m$, provided $m \ge n$.

\begin{lemma} \label{l.1} Let the function $a_m$ be as above and let us 
assume that $C^{-1} 2^{k/2} \le N_k \le C \, 2^{k/2}$. Then
$$ \frac{C^{-1} \, 2^{1/2}}{|I|-2^{-n}} \, 
   2^{-n} \frac{1 - 2^{-n(q-1/2)}}{1 - 2^{-(q-1/2)}} 
     \le M_q(a_m(\cdot),2^{-n}) \le 
   \frac{C \, 2^{-(q-1/2)}}{|I|-2^{-n}} \, 
   2^{-n} \frac{1 - 2^{-n(q-1/2)}}{1 - 2^{-(q-1/2)}} .$$
\end{lemma}

\proof In the same way as we did it for the inverse of the standard
Devil's staircase we can estimate $M_q(a_m(\cdot),2^{-n})$ as follows\,: 
\begin{eqnarray}
M_q(a_m(\cdot),2^{-n}) &:=&
\frac1{|I|-2^{-n}} \int_{I-(2^{-n})} [a_m(x+2^{-n}) - a_m(x)]^q \, dx
\nonumber\\
 &\le& \frac{2^{-n}}{|I|-2^{-n}} \sum_{k=0}^{n-1} N_{k+1} \, (h_{k+1}
+ 2^{-n})^q \nonumber\\
& \le& \frac{2^{-n} C}{|I|-2^{-n}} \sum_{k=0}^{n-1}
2^{(k+1)/2} (2^{-(k+1)} + 2^{-n})^q \nonumber \\
& =& \frac{C \, 2^{-n}
2^{-(q-1/2)}}{|I|-2^{-n}} \sum_{k=0}^{n-1} 2^{-k(q-1/2)} (1 + R_k) ,
\label{mbl5}
\end{eqnarray}
where
\begin{equation}
R_k := (1 + 2^{-n+k+1})^q - 1 \le \cases{ 2q 2^{-n+k+1},
&if $q>1$; \cr 2^{-n+k+1}, &if $q\le1$; \cr} .
\label{mbl6}
\end{equation} 
In (\ref{mbl5}), the term involving $R_k$ becomes negligible compared
to the other one as $n \to \infty$ and we are left with \,:
$$\frac{C \,
2^{-(q-1/2)}}{|I|-2^{-n}} \, 2^{-n} \frac{1 - 2^{-n(q-1/2)}}{1 -
2^{-(q-1/2)}} .$$ 
The estimate from below is obtained similarly. \qed

\begin{remark} To calculate $M_q(a(\cdot),l)$ for a random function $a(x)$ notice, 
that our estimates depend on $N_k$ linearly. Therefore inequalities
for the mathematical expectation of $N_k$ are enough to prove the
statement of Lemma~\ref{l.1} for the mathematical expectation of
$M_q(a(\cdot),l)$.
\end{remark}


\subsection{Estimate of $N_k$.}  \def\Pr{{\cal P}} \def\C{{\cal C}}
\label{s:rig2}

Let $\xi_\omega(x)$ be a realization of the standard Brownian motion 
on some finite interval $I$, and let $\C_w$ be the convex hull of the 
realization $w(y):=\int_o^y (\xi_\omega(x) + x) \, dx$. We denote by 
$S$ the set of end points of straight segments (corresponding to shocks) 
of $\C_w$. This set is a closed set of zero Lebesgue measure. 
For any fixed integer $k\ge0$, we introduce the following notation\,:
\begin{equation}
N_k := \#\{\Delta_j: \, 2^{-k-1} \le |\Delta_j| < 2^{-k}\},
\label{mbl7}
\end{equation}
where $\cup_j \Delta_j = I - S$ and $\Delta_j$ are intervals, whose 
end points lies in the set $S$.

Let us fix two points $x_1,x_2 \in I$ and a small number $0<\delta \ll 1$.
Consider two intervals 
$I_1:=[x_1 - |x_2-x_1|\delta, x_1 + |x_2-x_1|\delta)$ and 
$I_2:=[x_2 - |x_2-x_1|\delta, x_2 + |x_2-x_1|\delta)$.
We shall denote by $\zeta(I_1,I_2)$ the indicator function of the event 
that $\C_w$ has a straight segment whose endpoints lie inside the 
intervals $I_1,I_2$, respectively.

\begin{lemma} [Sinai] \label{l.N.1} There is a constant $C_0=C_0(\delta)>0$ 
such that $$ C_0^{-1} |x_2-x_1|^{1/2} \le \langle\zeta(I_1,I_2)\rangle
\le C_0 |x_2-x_1|^{1/2}. $$
\end{lemma}

\begin{lemma} \label{l.N.2} There is a constant $C>0$ such that 
$C^{-1} 2^{k/2} \le \langle N_k\rangle \le C \, 2^{k/2}$ for all $k$
large enough.
\end{lemma}

\proof Fix $k$ and choose a sufficiently large $m$. Decompose the segment
$I$ onto equal intervals $I_j:=[c_j^-,c_j^+)$, indexed from the left to the 
right, of length $2^{-k}/m$ and consider the pairs $I_i,I_j$ such that 
$m/2 \le j-i \le m+2$. Then
\begin{equation}
2^{-k-1} \le (j-i) 2^{-k}/m \le 2^{-k} - 2 \cdot 2^{-k}/m ,
\label{mbl8}
\end{equation}
which means that for any pair of points $x \in I_i$ and $y \in I_j$ 
we have $2^{-k-1} \le |x-y| < 2^{-k}$. Therefore,
\begin{equation}
N_k = \sum_{ij} \zeta(I_i,I_j) .
\label{mbl9}
\end{equation}
By Lemma~\ref{l.N.1} the mathematical expectation of $\zeta$ can be 
estimated as follows
\begin{equation}
C_0^{-1} \sqrt{1/2} \; 2^{-k/2} \le \langle \zeta(I_i,I_j)\rangle
   \le C_0 \; \sqrt{1-2/m} \; 2^{-k/2}.
\label{mbl10}
\end{equation}
Thus, summing over all pairs of indices, we have\,:
\begin{equation}
C^{-1} \; 2^{k/2} \le \langle N_k\rangle \le C_1 \; m^2 \; 2^k \; 2^{-k/2}
                             = C \, 2^{k/2} ,
\label{mbl11}
\end{equation}
which completes the proof. \qed

\section{Simulations and spurious scaling r\'egime}
\label{s:simulations}

The numerical strategy for solving Burgers equation with fractional Brownian
initial velocity has been described in detail in Ref.~\cite{verg}
(Section~5). Let us just recall some key points here.  In the
simulations, it is necessary to introduce both a large-scale and a
small-scale cutoff.  We find it convenient to work with periodic
velocity fields and to set the spatial period to unity. Then, the
scales accessible are clearly restricted to the range $\epsilon <
l<1$, where $\epsilon = 1/N$ is the inverse of the number of grid
points in the simulation. In Ref.~\cite{verg} up to $2^{20}$ grid
points were used. For reasons which will become clear, we had here to
work with even higher resolution, using $N=2^{22}$. Moments were
calculated by averaging over space and over about $1200$ realizations.
The constant $C$ appearing in (\ref{structfunc}) is always taken unity.

Fig.~\ref{f:mqhi} gives log-log plots of the moments $M_q(l)$ for
$h=1/2$ and 17 values of the exponent $q$ varying linearly between
zero and two. The graphs appear to all have the same unity slope at
small $l$'s and a different $q$-dependent slope at large $l$'s. 
\begin{figure}
\iffigs
\else
 \drawing 100 10 {$M_q$ as a function of $l$}
\fi
\caption{Moments of order $q$, as labeled, of Lagrangian increment
{\em vs.} separation $l$ for Brownian initial velocity ($h=1/2$) at
$t=0.5$.  The simulations used periodic conditions of unit size with a
mesh $\epsilon=2^{-22}$. Notice the conspicuous, but spurious, scaling
with unity exponent at small separations.}
\label{f:mqhi}
\end{figure}
Fig.~\ref{f:lsehi.5} shows the logarithmic derivative of $M_q(l)$ with
respect to $l$, a measure of the local scaling exponent. It is seen
that below a separation $l$ of about $10^{-4}$ all the exponents
$\tau_q$ go to unity, while at very large separations they appear to
approach the value $\tau_q=q$ (at least for $q$'s up to one).
\begin{figure}
\iffigs
\else
 \drawing 100 10 {Local scaling exponent for $h=1/2$}
\fi
\caption{Local scaling exponent obtained as the logarithmic derivative
of $M_q(l)$ with respect to $l$. Same conditions as Fig.~\protect\ref{f:mqhi}.}
\label{f:lsehi.5}
\end{figure}
The latter result is an immediate consequence of the fact that, at
separations much larger than the coalescence length $l_c$, the Lagrangian
map is very close to the identity. Hence, $M_q(l)\simeq l^q$.

We shall now show that, when $0\le q\le q_\star = h$, the former
result $\tau_q = 1$ is a numerical {\em artifact} due to a spurious
discretization effect, affecting all separations such that $l < l_{\rm
sp} \sim t \epsilon^h $, where $\epsilon$ is the numerical mesh.
By (\ref{craxi}), for a given $l$, the dominant contribution to $M_q(l)$ 
should come from those shocks with a Lagrangian length 
$\Delta a = 2^{-n}$ such that, 
\begin{equation}
l \sim t\, 2^{-nh} \sim t\, (\Delta a)^h.
\label{eulerlag}
\end{equation}
Clearly, no shocks can be represented which have $\Delta a <
\epsilon$, where $\epsilon$ is the numerical mesh. This gives a
cutoff in 
$l$ at 
\begin{equation}
l_{\rm sp} \sim t \,\epsilon^h.
\label{cutoff}
\end{equation}
For $l < l_{\rm sp}$, the numerically  measured Lagrangian increment will
typically take only two values\,:
$\epsilon$ with probability $l/l_{\rm sp}$ and
zero with probability  $1- l/l_{\rm sp}$. Hence,
\begin{equation}
M_q(l) \sim \epsilon^q  l/l_{\rm sp},
\label{spurmoment}
\end{equation}
 which implies $\tau_q = 1$. This is the spurious scaling
announced.

Thus, the scaling r\'egime with $\tau_q = q/h$, discussed in the
theoretical part of this paper, can be
observed only at scales such that
\begin{equation}
l_{\rm sp} \sim t \epsilon^h \ll l \ll l_c \sim
t^{1/(1-h)},
\label{limited}
\end{equation}
where 
\begin{equation}
l_c \sim t^{1/(1-h)}.
\label{coalescence}
\end{equation}
is the coalescence length \cite{verg}.

In practice (\ref{limited}) is a strong constraint\,: if we also want
to avoid contaminations of $\tau_q$ due to finite box size, we should
take  $l_c\sim t^{1/(1-h)}$ significantly smaller than unity.
Hence, for $h=1/2$, the range of nonspurious separations defined by
(\ref{limited}) will be too small to be  visible unless we work
at extremely high resolutions (very small $\epsilon$'s).

Inspection of (\ref{limited}) reveals, however, another strategy\,: we
can increase $h$ and thereby push the spurious range to smaller
separations. Since the fractional Brownian properties of the velocity
disappear at $h=1$, we chose $h=3/4$ as a trade-off.
Fig.~\ref{f:lsehi.75} shows the same result as in
Fig.~\protect\ref{f:lsehi.5} but, now, for $h=3/4$. 
\begin{figure}
\iffigs
\else
 \drawing 100 10 {Local scaling exponent for $h=3/4$}
\fi
\caption{Same as Fig.~\protect\ref{f:lsehi.5}, but for $h=3/4$. Notice
that the region of spurious scaling $\tau_q=1$ has shifted to smaller
separations and that a region with $\tau_q=q/h$ for small $q$ becomes 
visible.}
\label{f:lsehi.75}
\end{figure}
We observe that the region of spurious scaling has now been pushed
below $l\approx 10^{-5}$ and that for $0<q<q_\star =h= 3/4$ a kind of
plateau near $\tau_q=q/h$ is seen at the smallest separations not
affected by spurious scaling. Measuring directly the exponents
$\tau_q$ in this range produces Fig.~\ref{f:expfq} where $\tau_q$ is
plotted {\em vs.} the exponent $q$, both for $h=1/2$ and $h=3/4$. The
comparison with our theoretical prediction of
Section~\ref{s:heuristic} (the thick straight lines) is now
satisfactory, the only remaining discrepancies being caused by
unavoidable finite size effects which soften the phase transition.
\begin{figure}
\iffigs
\else
 \drawing 100 10 {Phase transition}
\fi
\caption{Scaling exponent measured at the smallest separation not
affected by spurious scaling, plotted {\em vs.} the order $q$ for
$h=1/2$ and $h=3/4$, as labeled. The  piecewise linear graphs
represent theoretical predictions.}
\label{f:expfq}
\end{figure}
\vspace{2mm}
\par\noindent {\bf Acknowledgements.}
Part of this work was done while E.A. and U.F. were visiting the
Mittag-Leffler Institute whose hospitality is gratefully acknowledged.
This work was supported by RFBR-INTAS through grant 95-IN-RU-03 723,
by the Swedish Institute (S.I.S.), by the Swedish Natural Science
Research Council through grant S-FO-1778-302 (E.A.), by a grant from
the {\em Fondation des Treilles\/}, by DRET (94/2582) and by the
French Ministry of Higher Education (M.B.).

\end{document}